# Superradiant Cherenkov-Wakefield radiation as THz source for FEL facilities

Klaus Floettmann[a], Francois Lemery[a], Martin Dohlus[a], Michela Marx[a], Vasili Tsakanov[b], Mikayel Ivanian[b]

[a] Deutsches Elektronen-Synchrotron, Notkestraße 85, 22607 Hamburg, Germany
[b] CANDLE SRI, 31 Acharyan Str., 0040 Yerevan Armenia

**Abstract**

An electron beam passing through a tube which is lined on the inside with a dielectric layer will radiate energy in the THz range due to the interaction with the boundary. The resonant enhancement of certain frequencies is conditioned by structure parameters as tube radius and permittivity of the dielectric layer. In low loss structures narrow-band radiation is generated which can be coupled out by suitable antennas. For higher frequencies the coupling to the resistive outer metal layer becomes increasingly important. The losses in the outer layer prohibit to reach high frequencies with narrow-band conditions. Instead short broad-band pulses can be generated with still attractive power levels.

In the first section of the paper a general theory of the impedance of a two-layer structure is presented and the coupling to the outer resistive layer is discussed. Approximate relations for the radiated energy, power and pulse length for a set of structure parameters are derived and compared to numerical results in the following section. Finally first numerical result of the out-coupling of the radiation by means of a Vlasov antenna and estimates of the achieved beam quality are presented.

## Introduction

X-ray Free-Electron Lasers are the brightest, tunable sources of short X-ray pulses available for basic scientific research. In order to unfold the full scientific potential of FELs it is however mandatory to complement the X-ray sources with suitable pump sources for pump-probe type experiments. The selective (non-) linear excitation of matter by electromagnetic radiation in the sub-THz to THz range (few meV photon energies) enables to deposit energy into specific low frequency modes of the material and thus allows to control the impact of various degrees of freedom onto material properties (Dhillon 2017). The quest for suitable THz sources for pump-probe type experiments at X-FEL facilities is thus a major development goal.

User requirements for the envisaged pump-probe experiments cover a broad parameter range of THz beam properties, for example frequencies spanning from 0.1 THz up to 30 THz combined with pulse energies of 3 mJ at 0.1 THz down to 0.03 µJ at 30 THz (Zalden 2018). Here it is assumed, that a focused beam size with a diameter of the wavelength can be achieved at the sample, so that the pulse energy scales quadratically with the frequency and field gradients in excess of 100 MV/m or equivalent magnetic fields of ~0.3 T are reached. A nearly diffraction limited transverse beam quality is hence desirable. Moreover a suitable THz source has to be synchronized to the XFEL facility with a low temporal jitter (below 20 fs at 5 THz) and it has to be able to deliver THz pulses at the operational repetition rate of the XFEL, so that the full potential of the facility can be employed.

Electron beam based THz sources can in principle cope with the high power, repetition rate, and frequency requirements. The freely available spent beam after a SASE undulator is an attractive option for this purpose, because it has still a high quality and a high beam power. Moreover it is naturally synchronized to the X-ray pulses and can fulfill all repetition rate requirements. (For a detailed discussion on the compensation of path length differences between THz and X-ray pulses see (Tanikawa 2018) ).

At the XUV FEL FLASH, for example, a 9 period electromagnetic undulator with 40 cm period length is installed behind the SASE undulator (Borisov 2006, Morozov 2007). The generated THz radiation in the range of 1.5 – 30 THz is transported through a ~65 m long evacuated beam line to the experimental chamber where it meets the XUV pulse on the sample (Gensch 2008, Willner 2008). Six refocusing mirrors in combination with planar mirrors keep the beam size under control and direct the beam to the experiment. A variable delay line allows for adjusting the relative timing of pump and probe beam.

Following the design ideas of FLASH the installation of a special undulator for the generation of THz radiation behind a SASE undulator at XFEL-facilities is discussed in (Tanikawa 2018). However, the high beam energy of XFEL facilities (>10 GeV) requires a total undulator length of 10 m with peak fields of up to 7.3 T and a period length of 1 m to comply with the requested THz parameters. While such an undulator appears to be technically feasibly with state-of-the-art superconducting technology the cost and complexity of such a device is not attractive.

Another conceivable option is the installation of a separate accelerator near the experimental hutch, because THz radiation can be generated with conventional undulator parameters already at some 10 MeV beam energy. Also the exploitation of a SASE process in the THz range is possible at these low electron beam energies (Schneidmiller 2012, Vardanyan 2014). Compact accelerators based on advanced concepts like, e.g., plasma acceleration are however not yet able to deal with the beam quality, charge and repetition rate requirements and cost, size and complexity of the system, based on conventional or advanced accelerator technology, are in any case very significant.

In this paper we consider another option which has not yet been discussed in detail with respect to the broad user requirements of XFEL pump-probe experiments, i.e. the utilization of superradiant THz radiation which is created by electron beams passing through vacuum pipes which are coated on the inside with a layer of, e.g., a dielectric material (Lemery 2019). The radiation process is treated in the literature as Cherenkov radiation in media with boundary conditions, e.g. (Bolotovski 1962), but also in terms of wakefields, e.g. (Ng 1990), where the focus is stronger on the effect of the boundaries onto the electron beam (induced energy spread and transverse forces). Both, radiation and wakefield effects as the induced energy spread are firmly described by theoretical models and by numerical results and have been experimentally demonstrated, primarily in the frequencies range of 0.1 – 1 THz, see e.g. ref (Hüning 2002, Cook 2009, Piot 2011, Antipov 2013, Smirnov 2015).

At relativistic beam energies the radiation wavelength in these structures becomes independent of the beam energy and is conditioned only by parameters of the vacuum tube, making it possible to use the spent electron beam after a SASE undulator. Especially for the lower frequency range (<10 THz) this technology offers a simple and cost effective solution.

In the following section a generalized representation of the impedance of two-layer vacuum tubes is presented. The bandwidth of the radiation is determined by losses in the dielectric and the coupling to the outer metallic layer which limits the narrow-band characteristics to the lower frequency range. Based on a discussion of the radiation characteristics relations to estimate radiated energy, power and pulse length for a set of structure parameters are derived and compared to numerical results in the following section. Finally first numerical result concerning the out-coupling of the radiation by means of a Vlasov antenna and first estimates of the achieved beam quality are presented.

## Impedance of a two-layer tube

A two-layer tube consists of an outer conducting tube which is coated on the inside by a thin layer of an electromagnetic meta-material. The term meta-material is justified here because the layer can for example be a low conductivity metal (layer thickness smaller than the skin depth), or a dielectric material or a purely geometrical structure, like regular radial grooves in a metal (corrugated structure), or simply a statistical rough metal surface (Novokhatsky 1998), see Figure 1. All layers are represented by an effective permittivity of the material and an effective layer thickness. For relatively thick, smooth layers the permittivity is in general the permittivity of the bulk material, but for thin layers the permittivity may deviate from the permittivity of the bulk material and depend on details of the surface morphology.

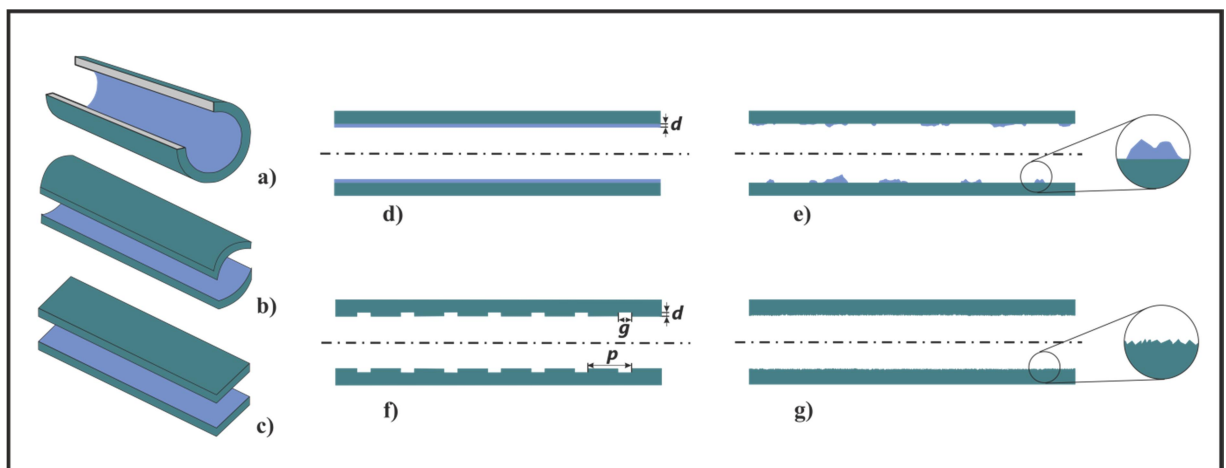

Figure 1: Possible radiator configurations are e. g. round tubes (a), and curved (b) or flat (c) parallel plate waveguides (PPWG). Parallel plate waveguides offer a tuning possibility via the distance of the plates but suffer from a reduced mode confinement as compared to the geometrical simple round tubes. While the outer structure of the radiator is a highly conducting metal, the inner coating can be a thin dielectric layer or a metal with low conductivity (d), but also patches of a dielectric material (e), regular geometrical structures like corrugations in a metal (f) or simply a rough metal surface (g). The last three cases are examples of meta-materials, because the radiation characteristics are described by an effective permittivity.

An example are corrugated structures with radial grooves of gap width *g*, depth *d* and period *p* in a metal base (Novokhaski 1997). The radiation characteristics of the structures are described by an effective permittivity $\varepsilon'$, related to the structure parameters by $\frac{\varepsilon'}{\varepsilon'-1} = \frac{p}{g}$ and the layer thickness *d*.

The simple geometry of round tubes (top left) simplifies the theoretical treatment of radiating structures. Besides round tubes offer a good mode confinement but they are limited in their tuning possibilities. Parallel-plate waveguides (bottom left) on the other hand can be tuned by variation of the plate distance, but the mode confinement is limited. Curved parallel plate waveguides improve the mode confinement and are tunable over a certain parameter range.

We present first a generalized description of the round tube impedance, the pure dielectric layer (dielectric lined) and the metallic layer (bimetallic) case are treated as limiting cases of this general form. For the radiation production low loss structures are preferable due to the lower bandwidth of the radiation. Small losses in the dielectric layer determine the bandwidth of the radiation in the low frequency range, while for higher frequencies the coupling to the outer metallic layer becomes relevant as will be discussed below.

A charged particle traveling through a (two-layer) tube interacts with the surrounding material through its space charge field, which is described in the laboratory frame by an infinite spectrum of waves traveling in radial direction. When a wave hits a material boundary it will be (partially) reflected, transmitted and/or diffracted. If the boundary is metallic the evanescent wave inside the metal will experience resistive loses which lead to a retarding force acting back onto the charged particle. Also in case of a dielectric boundary a retarding force acts onto the charged particle when the Cherenkov condition $\varepsilon'\beta^2 > 1$ is fulfilled for the partial wave in the dielectric (Schächter 1997). Here $\beta$ is the phase velocity of the wave which corresponds to the velocity of the charged particle.

In a two-layer tube the radiation fields interact also with the second boundary and the reflections on the interfaces lead to the general appearance of narrow-band resonances with typical frequencies in the lower THz range.

The fields near the axis of an arbitrary cylindrical symmetric structure can be expanded in terms of TE and TM, or as hybrid HEM modes. Field matching at the boundaries leads to solutions for two-layer (or general multi-layer) structures (Ivanyan 2008). The following discussion concentrates on the fundamental TM mode at relativistic energies, i.e. $\beta = 1$. In case that the second layer is treated as perfect metal with infinite conductivity the longitudinal impedance of an arbitrary nonmagnetic lining can be written as (Ivanyan 2004, Ivanyan 2008):

$$Z_\parallel(k) = i\frac{Z_0}{\pi k_z r_1^2}\left[1 + \frac{2\varepsilon_1}{r_1\varepsilon_0 k_r}\text{cth}(k_r d_1)\right]^{-1} \tag{1.1}$$

with the vacuum impedance $Z_0 = \sqrt{\mu_0/\varepsilon_0}$, the longitudinal propagation constant $k_z$ (which matches the free-space propagation constant $k_0$), and the transverse propagation constant in the first layer $k_r = k\sqrt{1-\varepsilon'-i\varepsilon''} = ik\sqrt{\varepsilon'-1+i\varepsilon''}$. Eq. (1.1) is valid in a high frequency range when $|k_r|r_1 \gg 1$.

The effective complex permittivity of the inner layer is described by $\varepsilon_1 = \varepsilon_0(\varepsilon' + i\varepsilon'')$, with the vacuum permittivity $\varepsilon_0$. While $\varepsilon'$ is a measure of the polarizability of the medium, $\varepsilon''$ describes losses in the material.

Losses in dielectric layers are small and often ignored; they limit however the shunt impedance and the quality factor of the resonance and are hence included in the derivation below. For a metal the permittivity is defined as $\varepsilon_1 = \varepsilon_0\left(1 + i\frac{\sigma_0}{\varepsilon_0 kc}\right)$, with the static conductivity $\sigma_0$, the wave number $k$ and the speed of light $c$. In contrast to the dielectric layer losses are dominant in metal layers. Moreover it is assumed that losses are independent of the frequency in dielectric layers, while they scale inversely with the frequency in metals. Despite these differences bimetallic and dielectric structures exhibit fundamentally similar resonance characteristics.

Expanding the cotangens hyperbolicus term in Eq. (1.1) to second order as: $\text{cth}(k_r d_1) = \frac{1}{k_r d_1} + \frac{k_r d_1}{3}$ allows to match Eq. (1.1) to the general impedance of a parallel resonance circuit (Ivanyan 2014):

$$Z_\parallel(k) = \frac{R}{1 + iQ\left(\frac{k_0^2 - k^2}{kk_0}\right)} \quad (1.2)$$

with the resonance wave number $k_0$ (associated to the resonance frequency $\nu_0$), the shunt impedance $R$ and the quality factor of the resonator $Q$ which is related to the bandwidth of the radiation by $\Delta\nu = \nu_0/Q$. Various asymptotic expression for the longitudinal monopole impedance of the two-layer dielectric tube are discussed in (Ivanyan 2019).

Table 1 summarizes relations resulting from the matching of Eqs. (1.1) and (1.2) for the general case and approximations for the dielectric and the bimetallic case (Ivanyan 2014). The necessary permittivity conditions are listed in the first row. For the resonance condition a thin layer, i.e. $2\varepsilon'd_1 \ll 3r_1$ is assumed. For thick layers the relations in the table tend to overestimate the resonance frequency. Parameter $A$ combines real and imaginary parts of the permittivity in a way suitable to derive the approximations. Parameter $\varsigma$ relates bandwidth $\Delta\nu$ and shunt impedance $R$ to the layer thickness $d_1$ and allows to optimize these parameters, as the bandwidth gets minimal and the shunt impedance maximal for $\varsigma = 1$.

The loss factor $K_\parallel$ describes the total energy loss per meter of a charged particle travelling through the structure. It is defined as integral over the complex impedance (Eq. (1.2)): $K_\parallel = -\frac{c}{2\pi}\int_{-\infty}^{\infty} Z(k)dk = -\pi R\Delta\nu$. Besides natural constants the loss factor depends only on the inner radius of the structure and is independent of the structure type.

Another important parameter listed in Table 1 is the group velocity (normalized to the speed of light) of the radiation pulse at the resonance frequency, which is not derived from the resonance equation Eq. (1.2) but follows the derivation in (Hüning 2002).

|  | **General** | **Dielectric** | **Bimetallic** |
|---|---|---|---|
| permittivity condition | $\varepsilon_1 = \varepsilon_0 (\varepsilon' + i\varepsilon'')$ | $\varepsilon'' \ll \varepsilon' - 1$ | $\varepsilon_1 = \varepsilon_0 \left(1 + i \frac{\sigma_0}{\varepsilon_0 kc}\right)$ |
| resonance condition $k_0^2$ | $\dfrac{2}{r_1 d_1} \dfrac{\varepsilon'(\varepsilon' - 1) + \varepsilon''^2}{(\varepsilon' - 1)^2 + \varepsilon''^2}$ | $\dfrac{2}{r_1 d_1} \dfrac{\varepsilon'}{\varepsilon' - 1}$ | $\dfrac{2}{r_1 d_1}$ |
| shunt impedance $R$ | $\dfrac{\sqrt{3} Z_0}{\pi r_1^2 d_1 k_0^2 \varepsilon'' A [\varsigma^{-1} + \varsigma]}$ | $\dfrac{\sqrt{3} Z_0}{2\pi r_1 [\varsigma^{-1} + \varsigma]} \dfrac{(\varepsilon' - 1)}{\varepsilon''}$ | $\dfrac{\sqrt{3} Z_0}{2\pi r_1 [\varsigma^{-1} + \varsigma]}$ |
| bandwidth $\Delta \nu$ | $\dfrac{d_1 k_0^2 \varepsilon'' c A [\varsigma^{-1} + \varsigma]}{2\sqrt{3}\pi}$ | $\dfrac{c [\varsigma^{-1} + \varsigma]}{\sqrt{3}\pi r_1} \dfrac{\varepsilon''}{\varepsilon' - 1}$ | $\dfrac{c [\varsigma^{-1} + \varsigma]}{\sqrt{3}\pi r_1}$ |
| parameter $A$ | $\dfrac{\sqrt{(\varepsilon' - 1)^2 + \varepsilon''^2}}{\varepsilon'(\varepsilon' - 1) + \varepsilon''^2}$ | $\dfrac{1}{\varepsilon'}$ | $\dfrac{\varepsilon_0 kc}{\sigma_0}$ |
| parameter $\varsigma$ | $\dfrac{d_1 k}{\sqrt{3}} \sqrt{(\varepsilon' - 1)^2 + \varepsilon''^2}$ | $\dfrac{d_1 k}{\sqrt{3}} (\varepsilon' - 1)$ | $\dfrac{d_1 \sigma_0 Z_0}{\sqrt{3}}$ |
| loss factor $K_\parallel$ | $-\dfrac{Z_0 c}{2\pi r_1^2}$ | $-\dfrac{Z_0 c}{2\pi r_1^2}$ | $-\dfrac{Z_0 c}{2\pi r_1^2}$ |
| group velocity $\beta_g$ |  | $1 - \dfrac{4 d_1}{r_1} \dfrac{(\varepsilon' - 1)}{\varepsilon'}$ | $1 - \dfrac{4 d_1}{r_1}$ |

Table 1: Resonance parameters for the general permittivity case and approximations for the dielectric and the bimetallic case. For details see text.

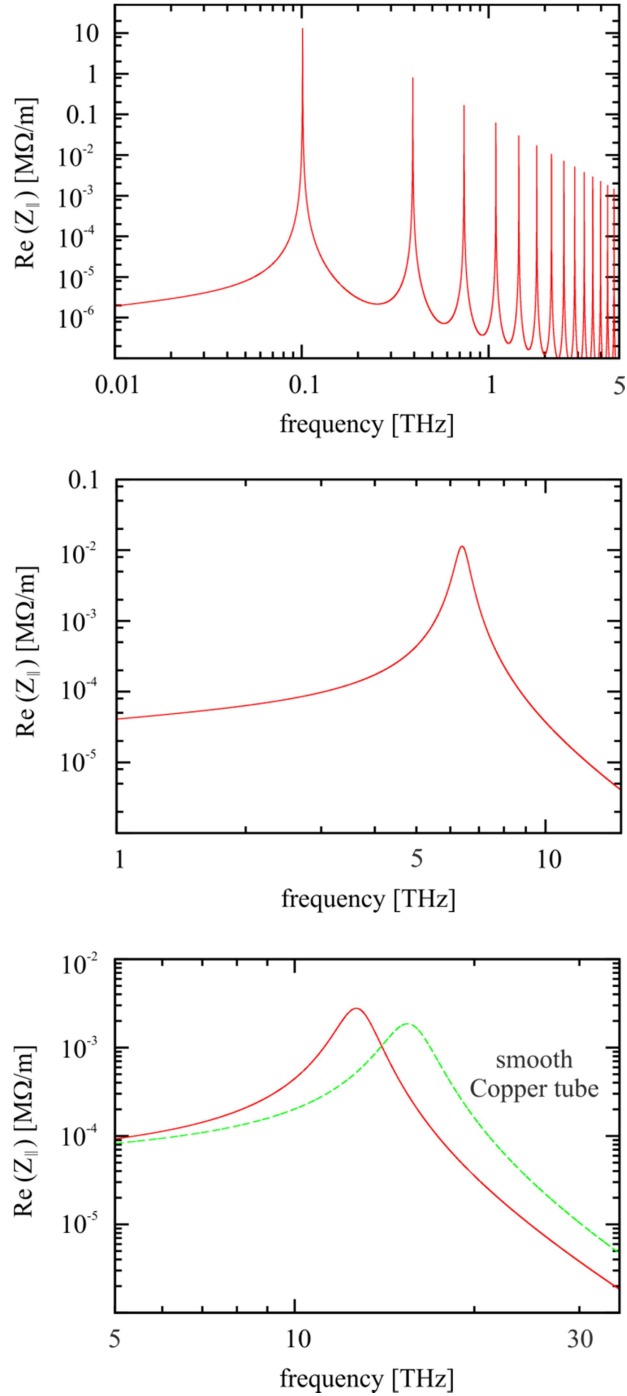

Figure 2: Real part of impedance vs. frequency for three different structures in double logarithmic representation. The first structure (top) is matched to 100 GHz. The inner radius $r_1$ is 2 mm, the permittivity of the dielectric layer $\varepsilon'$ is 4.41 (quartz), $\varepsilon''$ is assumed to be $4 \cdot 10^{-4}$ and the layer thickness $d_1$ is 225 µm. The second structure (middle) has the following parameters: radius $r_1 = 1$ mm, $\varepsilon' = 1.1$, $\varepsilon'' = 4 \cdot 10^{-4}$ and $d_1 = 1$ µm. It reaches a resonance frequency of 6.6 THz. For the third structure (bottom) it is finally assumed that the layer thickness $d_1$ is reduced to 0.1 µm, while all other parameters are identical to the second structure. The resonance frequency of the third structure is 15 THz. The outer layer is in all cases copper with a thickness larger than the skin depth and a conductivity $\sigma_0$ of 58.8 1/MΩm. For comparison the impedance of a pure resistive copper tube without surface roughness is included in the bottom plot.

The comparison of structures matched to three different frequencies in Figure 2 reveals details beyond the basic analytical description presented above. At rather thick dielectric layers (top, 100 GHz) many harmonics appear above the fundamental frequency. These resonances are caused by a periodic modulation of the transmission characteristics of the dielectric layer due to the etalon effect, i.e. the dielectric layer acts as a Fabry-Perot interferometer.

Mathematically the resonances are related to the periodicity of the hyperbolic cotangent (Eq. (1.1) in the complex plane, they coincide thus with the condition $\text{Im}(k_r d_1) = -n\pi$, where $n$ is an integer. The position of these resonance lines depends hence only on parameters of the dielectric layer and not on the radius of the tube. They are shifted to very high frequencies and low impedance values in case of thin layers and thus become irrelevant in the thin layer case. Moreover their dispersion curve levels off and finally does not cross the speed of line anymore. For a detailed discussion on the influence of the layer thickness on the dispersion curve see (Ivanyan 2020).

Structures matched to higher frequencies (middle, 6.6 THz and bottom, 15 THz) become increasingly more influenced by the finite conductivity of the outer metal layer.

In order to take the resistive loses in the outer layer into account Eq. (1.1) must be extended by a factor factor $\Gamma$ (Ivanyan 2011) as:

$$Z_\| = i \frac{Z_0}{\pi k_z r_1^2} \left[ 1 + \frac{2\varepsilon_1}{r_1 \varepsilon_0 k_r} \text{cth}(k_r d_1) \cdot \Gamma \right]^{-1} \tag{1.3}$$

$\Gamma$ is purely real and equal to 1 for the ideally conducting case, while it can be approximated by

$$\Gamma \cong 1 - \frac{1}{2F} - \frac{i}{2F}$$
$$F = \frac{(\varepsilon' - 1)}{\varepsilon'} d \sqrt{\frac{Z_0 \sigma_0 k_z}{2}} \tag{1.4}$$

when the resistivity of the outer layer has to be taken into account.

A resistive metal layer leads to a reduction of the resonance frequency and to an increased bandwidth.

The reduction of the resonance frequency is shown in Figure 3. The solid red line shows the resonance frequency when the resistivity of the outer copper layer ($\sigma_0$ = 58.8 1/MΩm) is taken into account in comparison to the expected resonance frequency following the resonance condition of Table 1 (indicated by the broken green line).

The resonance frequency of the coupled system follows the relation

$$k_{res}^2 = k_0^2 \left( 1 - \frac{1}{2} \frac{\varepsilon'}{(\varepsilon'-1)d} \sqrt{\frac{2}{Z_0 \sigma_0 k_{res}}} \right) \tag{1.5}$$

where $k_0$ is the resonance wavenumber of the tube with a layer of infinite conductivity (Table 1). Eq. (1.5) can be solved by an iterative procedure. The coupled resonance frequency converges to the peak value of the impedance of a pure metallic tube which is approximately given by $\nu_0 \approx \frac{c}{2\pi} \left( \frac{\sqrt{2Z_0 \sigma_0}}{r} \right)^{2/3} \approx 17$ THz for a copper tube with 1 mm radius. This frequency sets an upper limit for the resonance properties of a two-layer structure for this radius.

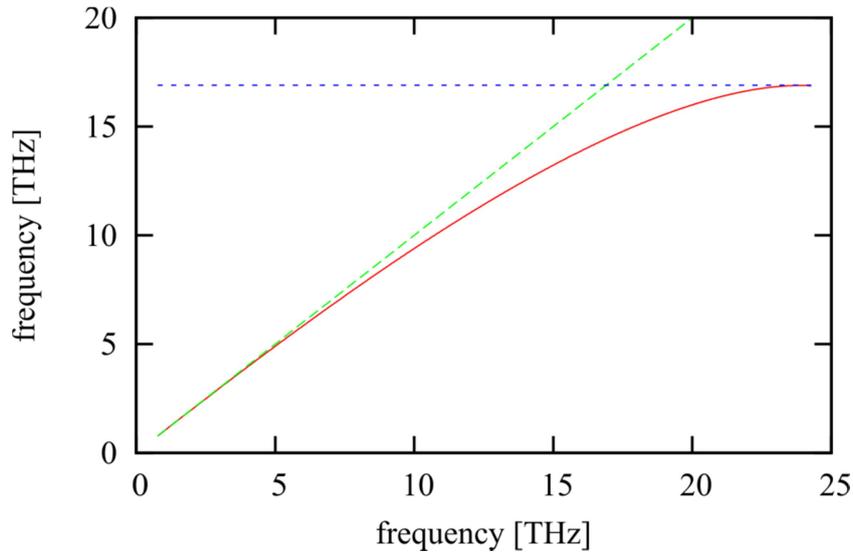

Figure 3: Resonance frequency reduction due to the resistivity of the outer copper layer. The green line represents the expectation according to the resonance condition of Table 1. For the red line the resistivity is taken into account (Eq. (1.5)). The blue line shows the peak position of a pure resistive metal tube.

The coupling to the resistive layer disturbs the resonance character of the radiation so that the impedance cannot be described by a resonator anymore when the influence of the metal becomes too strong, i.e. it develops more and more into a broadband radiation. Note that the loss factor, i.e. the total radiated energy does however not change due to the coupling to the resistive layer.

The analytical results presented in this section are based on a low frequency model of the metal permittivity which assumes a frequency independent conductivity. For higher frequencies (material dependent >10 THz) the frequency dependent conductivity following the Drude model of metals should be employed. (The Drude model is employed for the smooth copper tube impedance in Figure 2). Moreover the anomalous skin effect may become significant if for example the structure shall be cooled in order to increase the conductivity. In that sense the above results are approximate when higher frequencies are considered. More developments are required to improve our understanding of the high frequency limits.

# Radiation characteristics

The following discussion concentrates on the dielectric case as presented in Table 1. The focus is on the generation of short pulses, i.e. the frequency content of the pulses will be dominated by the $\frac{\sin(\Delta\omega\tau)}{\Delta\omega\tau}$ term related to the pulse length $\tau$. The natural bandwidth of the radiation can thus to first order be ignored and besides the resonance condition only the loss factor and the group velocity are required to estimate radiation parameters.

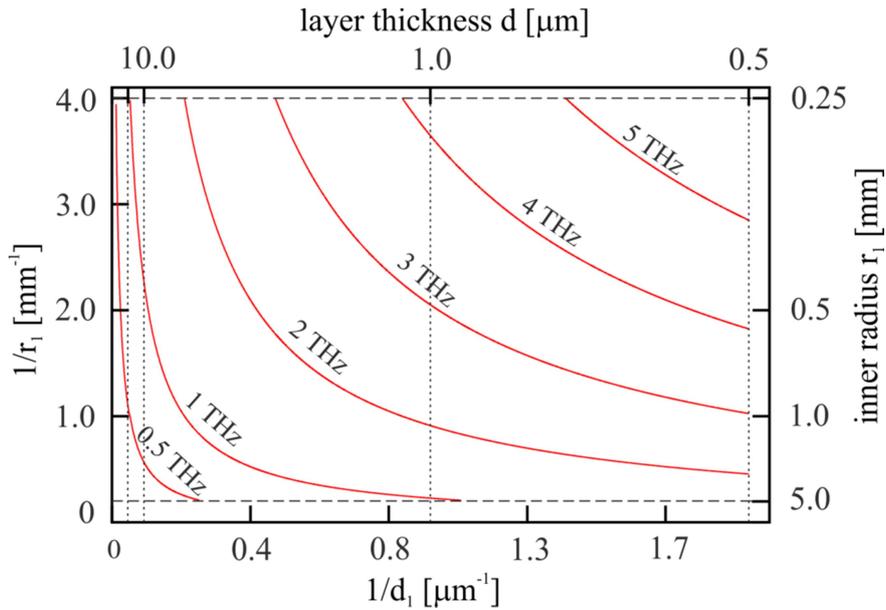

Figure 4: Resonance frequency as function of inner radius und layer thickness for a dielectric structure with $\varepsilon' \gg 1$. For a better separation of the frequency contour lines inverse scales are used.

Figure 4 shows the resonance frequency for a dielectric structure with large permittivity $\varepsilon' \gg 1$ as function of tube radius and layer thickness. In this plot the inner radius ranges from 0.25 to 5 mm, while for the thickness of the inner coating a minimum value of 0.5 µm is chosen.

The lower frequency range from a few hundred GHz up to a few THz can be easily covered by dielectric structures it is however unclear what the highest achievable frequency for this technology is. Note, that the frequencies scale with the ratio $\varepsilon'/(\varepsilon'-1)$. Thus for $\varepsilon' = 2$ the 5 THz line would scale up to 10 THz, while it would require $\varepsilon' = 1.2$ to push it up to 30 THz. Corrugated structures as sketched in Figure 1 reach permittivity's below 2 and appear hence to be suitable structures for low and medium frequencies. As discussed in the previous section the finite conductivity of the outer layer (or the base material in case of a corrugated structure) prohibits to reach high frequencies with narrow band conditions in two-layer structures. Options as cooling of the metallic tubes to increase the conductivity or more complex structures (multi-layer, photonic band gap) may open a way to higher frequencies but have not yet been investigated in detail. Also filtering of the broadband radiation of metal tubes is an option as the required power levels decrease with increasing frequency.

Relevant parameters for the high frequency operation, besides the permittivity, are the layer thickness and the radius of the structures.

For the case of negligible losses, i.e. an ideally conducting outer layer and $\varepsilon' = 0$ for the inner layer the impedance reduces to a delta function and the wake field is purely monochromatic with an amplitude given by the loss factor as:

$$W^\delta(s) = 2K_\| \cos(k_0 s) \tag{1.6}$$

Eq. (1.6) describes the field induced by a single electron as function of the longitudinal position $s$ behind and relative to the position of the electron. The field induced by a bunch is proportional to the bunch charge but also on the frequency contend of the charge distribution:

$$W(s) = 2qFK_\| \cos(k_0 s) \tag{1.7}$$

where $F = F(\nu_0)$ stands for the bunch form factor, i.e. the normalized Fourier component at the resonance frequency of the structure.

The total energy radiated in a structure of length $L$ follows thus as:

$$E^{rad} = q^2 F^2 |K_\|| L = q^2 F^2 L \frac{Z_0 c}{2\pi r_1^2} \tag{1.8}$$

The pulse length of the radiation pulse is related to the group velocity (Table 1) by the relation

$$\tau = \frac{L}{c}(1 - \beta_g) = L \frac{(\varepsilon' - 1)}{\varepsilon'} \frac{4d_1}{r_1 c} \tag{1.9}$$

From the resonance condition follows $\frac{\varepsilon' - 1}{\varepsilon'} = k_0^2 \frac{r_1 d_1}{2}$, which leads to

$$\tau = \frac{8}{k_0^2 r_1^2 c} L \tag{1.10}$$

Finally the power can be calculated as ratio of radiated energy Eq. (1.8) and pulse length Eq. (1.10) as:

$$P^{rad} = q^2 F^2 \frac{Z_0 k_0^2 c}{16\pi} \tag{1.11}$$

To estimate the tube length required to cope with the user requests as stated in (Zalden 2018) the bunch form factor is needed. This is shown in Figure 5 for the European XFEL operated at different bunch charges. In case of operation at lower bunch charge the beam needs to be stronger compressed to increase the bunch current to the level required for the FEL process and thus a high bunch form factor is reached at higher frequencies. Below ~1 THz the bunch form factor is close to 1 even for a bunch charge of 1 nC.

Based on this it would for example require a structure of only 66 cm length when the structure radius is 2 mm and the bunch charge is 1 nC to generate an energy of 3 mJ (Eq. (1.8)) at 0.1 THz (Zalden 2018). The temporal length of the radiation pulse would be 1 ns (100 cycles) (Eq. (1.10)). And the power reached in this case is 3 MW (Eq. (1.11)).

The resonance frequency does not enter Eq. (1.8), a structure with the same dimensions but different dielectric layer would thus generate the same energy up to a resonant frequency of about 1~THz, when the bunch form factor starts to decrease for the 1 nC case. The pulses at higher frequency are however shorter and the power higher as $k_0$ enters into Eq. (1.10) and Eq. (1.11). Since less energy is requested at higher frequencies (Zalden 2018) shorter structures or less charge would still be sufficient to cope with the user requirements.

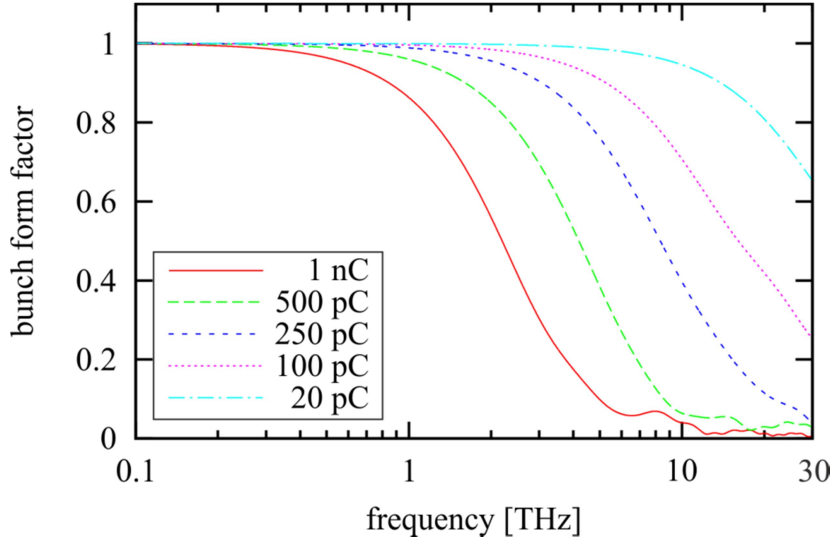

Figure 5: Bunch form factor for different bunch charges at the European XFEL. The particle distributions are generated by start to end simulations up to an energy of 17.5 GeV at the entrance of the undulators. The FEL interaction is thus not yet taken into account (Zagorodnov 2013).

Note that Eq. (1.8) and Eq. (1.10) scale as $L/r_1^2$, a larger radius requires thus a longer structure to generate the same amount of energy, but the pulse length and the power don't change if the length of the structure is adjusted accordingly.

At higher frequencies the bunch form factor shrinks considerably. Lower bunch charges achieve higher form factors so that it can become advantageous to reduce the bunch charge when radiation above ~3 THz shall be generated. On the other hand is the higher bunch charge of 1 nC preferable for the lower frequency range but it is not mandatory as longer structures and/or smaller radii are not excluded.

The reduced energy requirements still allow generating sufficient amounts of energy also at higher frequencies. At 6.6 THz (Figure 4, middle) only 0.7 µJ radiation energy are requested which requires less than a centimeter structure length (1 mm radius) at charges between 1 nC and 100 pC.

In view of the discussion of the influence of the finite conductivity of the metallic layer it is clear that the approximate relation (Eqs (1.8) - (1.11)) are not applicable at higher frequencies. It is nevertheless interesting to note, that only short structures (less than a centimeter) would be required to fulfill the user requests. This allows considering very small radii of the structures.

In Figure 6 THz wave forms generated by electron bunches for the impedances introduced in Figure 2 are compared. The traces are numerically calculated by an inverse Fourier transform of the complex impedance

$$W(t) = \int_{-\infty}^{\infty} Z_{\parallel} e^{-i2\pi \nu t} dt \qquad (1.12)$$

Neither dispersion nor transition effects are included and the time axis corresponds to a wave traveling in vacuum. (The wavelength is reduced inside of the structure due to the lower group velocity.)

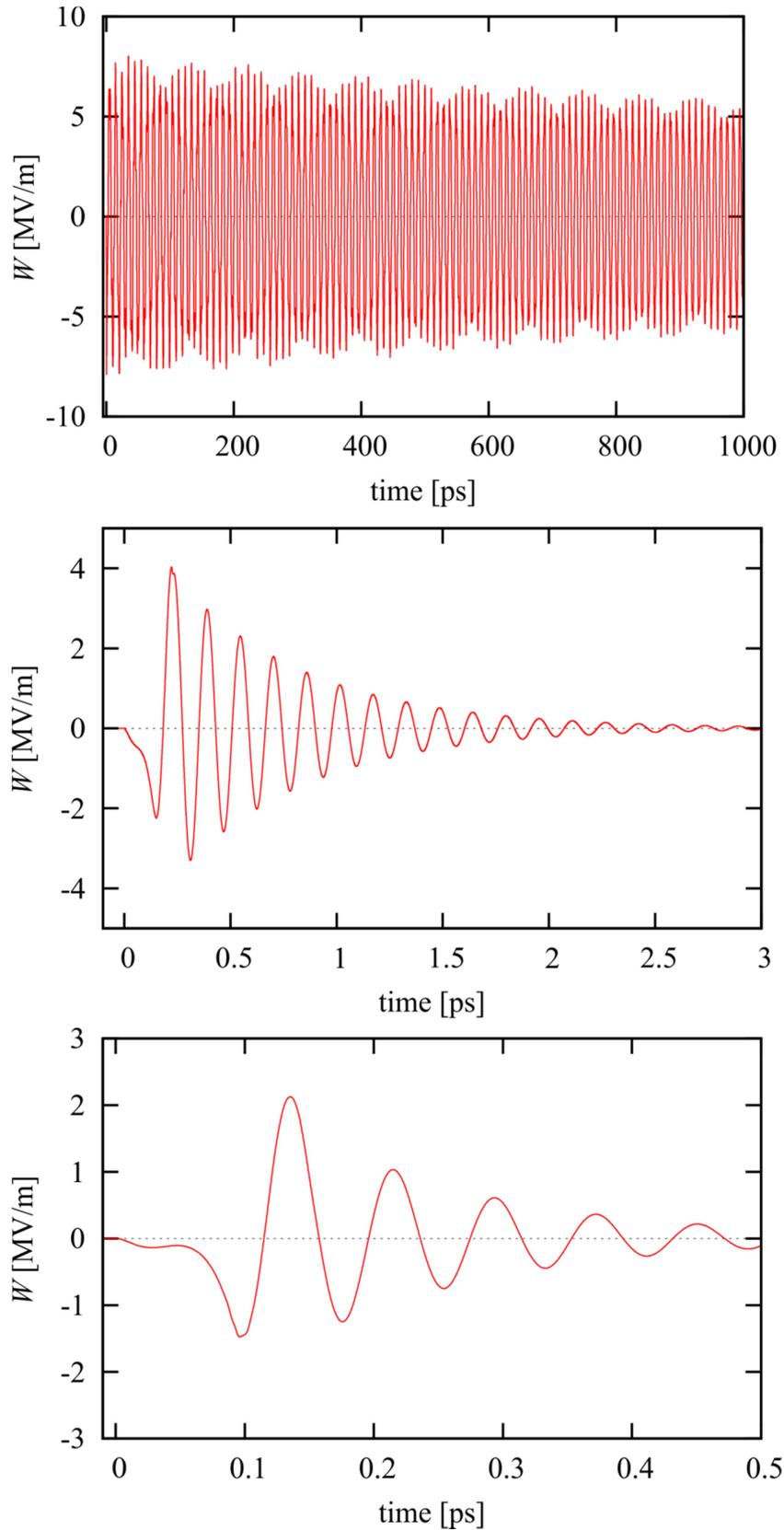

Figure 6: THz wave forms generated by electron beams in dielectric tubes. The parameters of the tubes correspond to the impedance plots shown in Figure 4. For the top plot (0.1 THz) a bunch charge of 1 nC is assumed. The pulse amplitude is modulated by the higher harmonics of the structure which are included in this numerical result. The pulse length of 1000 ps corresponds to 66 cm structure length. At 6.6 THz (middle) the pulse is strongly damped and decays with an exponential decay time of about 0.6 ps. Here a bunch charge of 0.5 nC is assumed. The bottom plot shows the 15 THz case with a charge of 0.25 nC. The pulse decays with about 150 fs decay time.

At higher frequencies the damping is quite strong, so that a steady state is reached in very short structures. At low frequencies the damping can be very small, so that long nearly sinusoidal waves can be generated. Of course it is also possible to increase the damping for the lower frequency range if a larger bandwidth is desired.
The considerations above aim to serve as a guideline only. A full design requires numerical simulations including transition effects at the entrance and the exit of the structure. An overhead to account for transport losses has to be included and possible modifications of the pulse structure due to the transport line need to be considered. In the following section briefly some aspects of the out coupling of the THz radiation and some problems concerning the numerical simulation are presented.

## Out-coupling and transverse beam quality

An important aspect of the THz generation with dielectric tubes is the out-coupling of the radiation from the generating tube into a free space transport system. Experimentally two concepts have been tested. A straight line extraction of the radiation can be realized by means of tapered structures ending in horn antennas (Cook 2009, Smirnov 2015). Since the radiation travels on a straight path overlapping the electron beam (and potentially the FEL radiation) the THz radiation needs to be separated from the electron beam, e.g., by a mirror with a hole. A disadvantage of this approach is that the mirror distorts the wave front of the THz pulse due to the hole. Moreover diffraction radiation is generated by the electron beam passing through the hole which adds to the THz pulse. If the FEL beam is separated from the electron beam before it enters the THz structure it would of course be possible to employ a dipole magnet for the separation of THz and electron beam.
A second approach is a so-called Vlasov antenna (Vlasov 1974, Antipov 2016). Here the tube is cut at an angle which transforms the radiation field into a highly directed beam at an angle relative to the beam axis as shown in Figure 7. In the simulations performed with the CST-MWS frequency domain solver for a frequency of 300 GHz a high transmission above 0.95% of the power was realized at different cut angles (tolerances appear to be uncritical). To characterize the radiation properties the field has been transformed from the finite calculation domain into the far zone, and the coordinate system has been rotated so that the $z$-axis points into the direction of maximal power flux. This corresponds to a counter clockwise rotation of about 30 deg around the $y$-axis, that points out of the plane of the diagram.

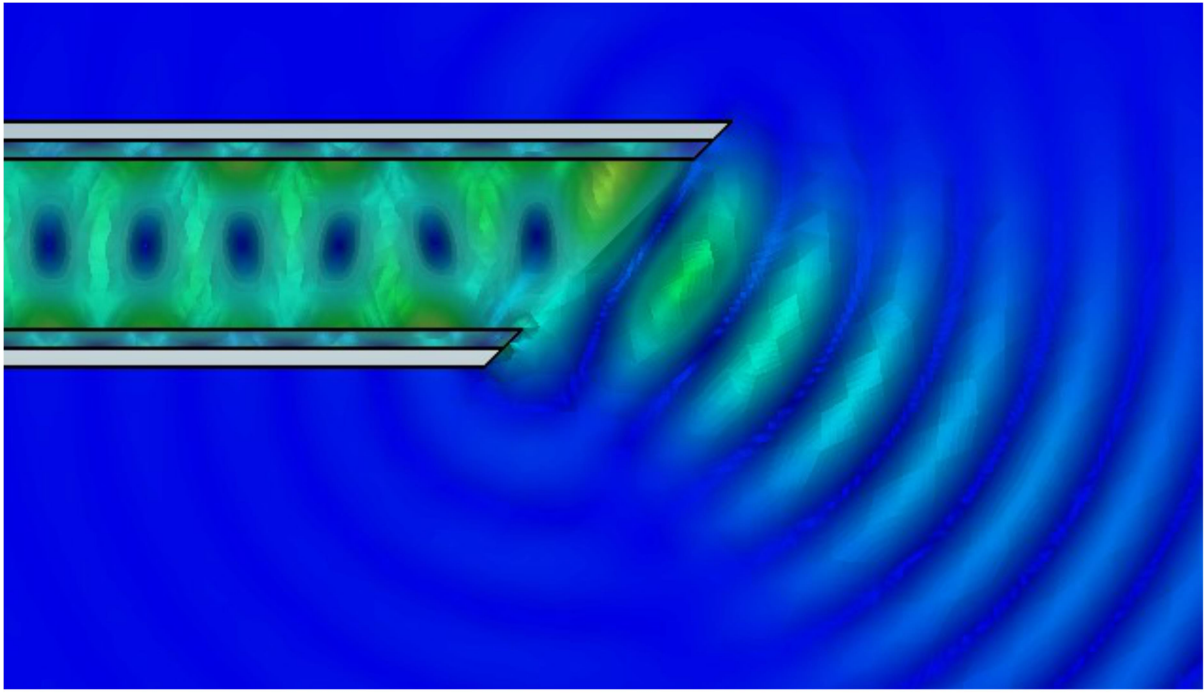
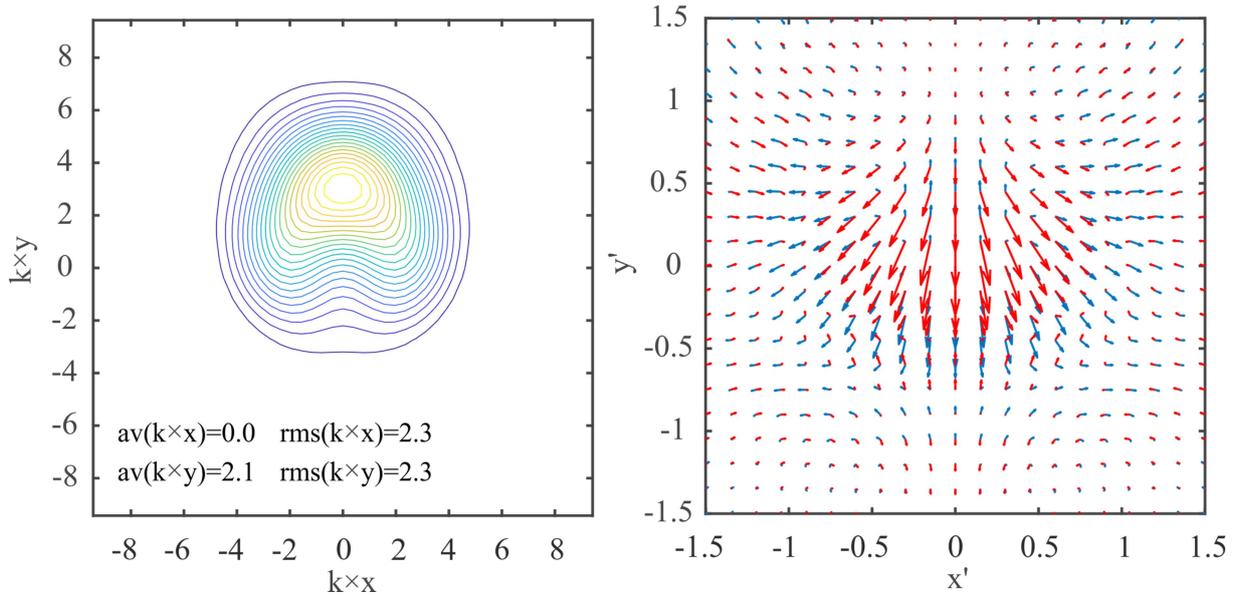

Figure 7: Simulation of a Vlasov antenna with 300 GHz central frequency. The tube has a radius of 450 µm, a layer thickness of 100 µm, and $\varepsilon' = 3.8$ (fused silica). The cut angle has been varied between 25 and 65 degree. Lower left: Perpendicular to the plane component of the Poynting flux through a plane perpendicular to forward direction. The Poynting flux is calculated by a far-field to near-field transformation into the plane with minimal rms dimensions of the pattern; $k$ is the wavenumber. Lower right: Vector components of the electric field in the far zone for $\mathbf{e}_r = \dfrac{x' \sin r}{r'} \mathbf{e}_x + \dfrac{y' \sin r}{r'} \mathbf{e}_y + \cos r' \mathbf{e}_z$ with $r' = \sqrt{x'^2 + y'^2}$. The real and imaginary parts are shown in blue and red.

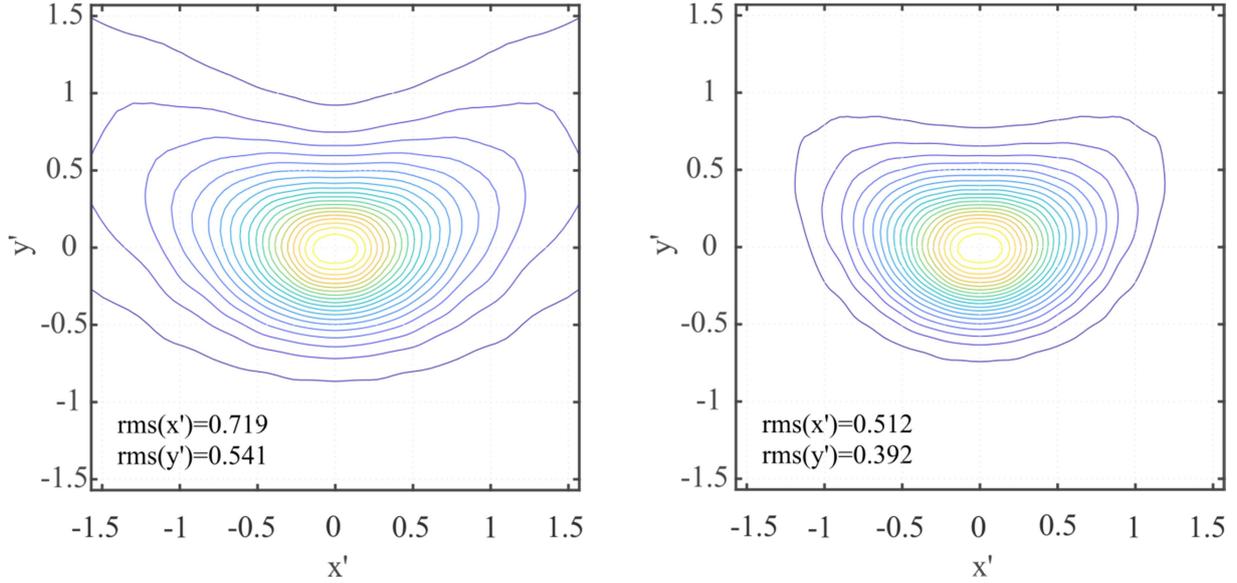

Figure 8: Poynting flux in far zone. Plotted is the radial component into direction $\mathbf{e}_r \propto x'\mathbf{e}_x + y'\mathbf{e}_y + \mathbf{e}_z$ (left) and into direction $\mathbf{e}_r = \frac{x'\sin r}{r'}\mathbf{e}_x + \frac{y'\sin r}{r'}\mathbf{e}_y + \cos r'\mathbf{e}_z$ with $r' = \sqrt{x'^2 + y'^2}$ (right).

Figure 8 presents contour plots of the Poynting-flux through a far sphere with different representations of the directional vector $\mathbf{e}_r$ as function of angular coordinates $x'$ and $y'$. For Figure 8, left these coordinates have the meaning of the slope in x- and y-direction as it is convention for beam dynamics. For non-narrow characteristics it is appropriate to use a spherical description. In Figure 8, right the spherical angles $\varphi$ and $\theta$ are the angle and length of $x + iy$. (The difference between both diagrams increases with the distance from the origin.) The polarization of the wave depends on its direction and has even a circular component, see Figure 7 (lower right).

The Poynting flux in Figure 7 (lower left) is obtained by a far- to near-field transformation to a plane $z$ = const., here the plane with minimal rms dimensions. The beam quality can be estimated from the rms dimensions in the far (characteristic) and near (waist of the back-propagated beam). It is about twice the diffraction limited beam quality. Further studies at different frequencies and optimized geometries as well as experimental verifications are required.

## Conclusion & Outlook

We present a generalized treatment of the impedance of a two-layer structure in the form of a matched resonance circuit. Losses in the dielectric layer and especially in the coupled outer metal layer determine frequency content and bandwidth of the radiation. At higher frequencies (very thin dielectric layer) the impedance approaches the typical resistive wakefield impedance of a metal tube. The narrow-band characteristics of the lower frequency range can thus not be maintained anymore.

Superradiant Cherenkov-wakefield radiation presents an attractive option as THz source for FEL facilities especially for the lower frequency range up to a few THz. The user requirements can be met with moderate geometrical parameters of the tubes. At higher frequencies less energy is requested, so that shorter tubes can be employed which allows to reduce the radius further. In addition very thin dielectric layers are required to push this technology to higher frequencies.

While robust structures with relatively thick dielectric layers can be made for example from quartz tubes which are coated with copper on the outside, the production of thin layers can be based, e.g., on sputtering or simple oxidation of a metal tube on the inside. Corrugated structures offer another way to reach a small effective permittivity and a thin layer thickness.

The structures, especially the all-metal structures, are in general robust and will withstand some particle loss and energy deposition (heat) which may appear during operation. A first study of a corrugated structure for LCLS II (Bane 2017) indicates manageable heat load problems for the considered parameters. Operational experience with a high power beam, as for example the beam of the European XFEL, is however missing at this point in time. Also the experimental investigation of high field effects which might limit the performance requires more attention (O`Shea 2019).

Round tube structures are attractive for their radiation characteristics, but their tuning capability is limited. It is however conceivable to install a larger number of tubes on a movable stage, so that not only a substantial frequency range can be covered, also tubes for different pulse lengths or bandwidths can be provided. This flexibility is unique to this technique.

Curved parallel plate waveguides are another radiator option with increased tuning capability. These structures can potentially also produce variable chirped pulses for later THz-pulse compression. Also the generation of two or more colors in a single pulse by passing the beam through structures with different radii is conceivable. Cherenkov-wakefield radiators are hence not only simple and inexpensive in comparison to other options for the generation of THz radiation for pump-probe experiments at FEL facilities. They also offer a high flexibility over a broad range of parameters to cope with various user requests.

## References

Antipov S. et al. (2016) 'Efficient extraction of high power THz radiation generated by an ultra-relativistic electron beam in a dielectric loaded waveguide', Appl. Phys. Lett 109, 142901.
http://dx.doi.org/10.1063/1.4963762

Antipov, S. et al. (2013) 'High power terahertz radiation source based on electron beam wakefields', Rev. Sci. Instr. 84, 022706.
https://doi.org/10.1103/PhysRevLett.111.134802

Bane, K., Stupakov, G., Gjonaj, E. (2017) 'Joule heating in a flat dechirper', Phys. Rev. Accel. Beams 20, 054403.
https://doi.org/10.1103/PhysRevAccelBeams.20.054403

Bolotovski, B. M. (1962) 'Theory of Cherenkov Radiation (III)', Soviet Physics Uspekhi 4, 781,
http://dx.doi.org/10.1070/PU1962v004n05ABEH003380
Rus org.: Usp. Fiz. Nauk 75, 295, (1961)

Borisov, V. et al. (2006) 'Simulation of Electromagnetic Undulator for Far Infrared Coherent Source of TTF at DESY', Proc. of Euro. Part. Accel. Conf., Edinburgh.
http://accelconf.web.cern.ch/AccelConf/e06/PAPERS/THPLS133.PDF

Cook, M. et al. (2009) 'Observation of Narrow-Band Terahertz Coherent Cherenkov Radiation from a Cylindrical Dielectric-Lined Waveguide', Physical Review Letters 103, 095003.
https://doi.org/10.1103/PhysRevLett.103.095003

Dhillon, S. S. et al. (2017) 'The 2017 terahertz science and technology roadmap', J. Phys. D: Appl. Phys. 50 043001.
https://doi.org/10.1088/1361-6463/50/4/043001

Gensch, M. et al. (2008) 'New Infrared Undulator Beamline at FLASH', Infrared & Physics Technology, 51, 5.
https://doi.org/10.1016/j.infrared.2007.12.032

Hüning, M. (2002) 'Analysis of Surface Roughness Wake Fields and Longitudinal Phase Space in a Linear Electron Accelerator', DESY-Thesis-2002-029.
http://www-library.desy.de/cgi-bin/showprep.pl?desy-thesis-02-029

Hüning, M., Schlarb, H., Schmüser, P., Timm, M. (2002) 'Measurements of Harmonic Wake Fields Excited by Rough Surfaces', Phys. Rev. Lett. 88, 074802.
https://doi.org/10.1103/PhysRevLett.88.074802

Ivanyan, M., Tsakanov, V. (2004) 'Longitudinal impedance of two-layer tube', Phys. Rev. ST Accel. Beams 7, 114402.
https://doi.org/10.1103/PhysRevSTAB.7.114402

Ivanyan, M., Laziev, E., Tsakanov, V., Vardanyan, A., Heifets, S., Tsakanian, A. (2008) 'Multilayer tube impedance and external radiation', Phys. Rev. ST Accel. Beams 11, 084001.
https://doi.org/10.1103/PhysRevSTAB.11.084001

Ivanyan, M., Tsakanov, V. (2011) 'Coupling Impedance of Rough Resistive Pipe', Proc. of Int. Part. Accel. Conf., San Sebastian.
http://accelconf.web.cern.ch/AccelConf/IPAC2011/papers/mops045.pdf

Ivanyan, M., Grigoryan, A., Tsakanian, A., Tsakanov, V. (2014) 'Narrow-band impedance of a round metallic pipe with a low conductive thin layer', Phys. Rev. ST Accel. Beams 17, 021302.
https://doi.org/10.1103/PhysRevSTAB.17.021302

Ivanyan, M., Aslyan, L., Tsakanov, V., Lemery, F., Floettmann, K. (2019) 'On the Resonant Properties of the Metal-Dielectric Waveguide Impedance', Proc. of 4th EAAC workshop, Elba, accepted for publication.

Ivanyan, M., Aslyan, L., Tsakanov, V., Lemery, F., Floettmann, K. (2020) 'Wake fields in conducting waveguides with a lossy dielectric channel', Phys. Rev. Accel. Beams 23, 041301.
https://doi.org/10.1103/PhysRevAccelBeams.23.041301

Lemery, F., Floettmann, K., Dohlus, M., Marx, M., Ivanyan, M., Tsakanov, V. (2019) 'A versatile THz source for high-repetition rate XFELs', Proc. of Free-Electron Laser Conf., Hamburg.
http://accelconf.web.cern.ch/AccelConf/fel2019/papers/thp049.pdf

Morozov, N. et al. (2007) 'Magnetic Measurements of the FLASH Infrared Undulator', Proc. of Free-Electron Laser Conf. 2007, Novosibirsk.
http://accelconf.web.cern.ch/AccelConf/f07/PAPERS/WEPPH003.PDF

Ng, K.-Y. (1990) 'Wake fields in a dielectric-lined waveguide', Physical Review D 42, 5.
https://doi.org/10.1103/PhysRevD.42.1819

Novokhaski, A., Mosnier, A. (1997) 'Wakefields of short Bunches in the Canal covered with thin dielectric layer', Proc. of Part. Accel. Conf., Vancouver.
http://accelconf.web.cern.ch/accelconf/pac97/papers/pdf/5V028.PDF

Novokhatsky, A., Timm, M., Weiland, T. (1998) 'The surface roughness wake field effect', Proc. of Intl. Comp. Accel. Phys. Conf., Monterey.
http://www.slac.stanford.edu/xorg/icap98/

O`Shea, B. D., Andonian, G., Barber, S. K., Clarke, C. I., Hoang, P. D., Hogan, M. J., Naranjo, B., Williams, O. B., Yakimenko, V., Rosenzweig, J. B. (2019) 'Conductivity Induced by High-Field Terahertz Waves in Dielectric Material', Phys. Rev. Lett. 123, 134801.
https://doi.org/10.1103/PhysRevLett.123.134801

Piot, P. et al. (2011) 'Observation of Coherently-Enhanced Tunable Narrow-Band Terahertz Transition Radiation from a Relativistic Sub-Picosecond Electron Bunch Train', Appl. Phys. Lett. 98, 261501.
https://doi.org/10.1063/1.3604017

Schächter, L., Schieber, D. (1997) 'On the characteristics of the Cherenkov and Ohm forces', Nucl. Instrum. Methods Phys. Res. Sect. A 388, 8.
https://doi.org/10.1016/S0168-9002(97)00311-2

Smirnov, V. et al. (2015) 'Observation of a variable sub-THz radiation driven by a low energy electron beam from a thermionic rf electron gun', Phys. Rev. ST Accel. Beams 18, 090703.
https://doi.org/10.1103/PhysRevSTAB.18.090703

Schneidmiller, E.A., Yurkov, M.V., Krasilnikov, M., Stephan, F. (2012) 'Tunable IR/THz source for pump probe experiments at the European XFEL', Proc. of Free-Electron Laser Conf., Nara.
http://accelconf.web.cern.ch/AccelConf/FEL2012/papers/wepd55.pdf

Tanikawa, T. et al. (2018) 'Superradiant Undulator Radiation for Selective THz Control Experiments at XFELs', European XFEL Report XFEL.EU TN-2018-002.
https://xfel.tind.io/record/1563

Vardanyan, T. et al. (2014) 'ALPHA – The THz Radiation Source Based on AREAL', Proc. of Free-Electron Laser Conf., Basel.
http://accelconf.web.cern.ch/AccelConf/FEL2014/papers/tup083.pdf

Vlasov, S. N., Orlova, I. M. (1974) 'Quasioptical transformer which transforms the waves in a waveguide having a circular cross section into a highly directional wave beam' Radiophys. Quantum Electron. 17(1), 115–119.
https://link.springer.com/article/10.1007%2FBF01037072

Willner, A. (2008) 'Investigations into the FLASH Infrared Undulator as an Electron Beam Diagnostic Tool', Diploma Thesis, TESLA-FEL 2008-04.
https://flash.desy.de/sites2009/site_vuvfel/content/e403/e1642/e2308/e2310/infoboxContent2325/TESLA-FEL2008-04.pdf

Zagorodnov, I. (2013) Homepage of the XFEL beam Dynamics group.
http://www.desy.de/xfel-beam/s2e/xfel.html

Zalden, P. et al. (2018) 'Terahertz Science at European XFEL', European XFEL Report XFEL.EU TN-2018-001-01.0.
https://xfel.tind.io/record/1564